\documentclass[preprint2,times,tighten]{aastex6}
\usepackage{amsmath,amstext}
\usepackage[figure,figure*]{hypcap}
\usepackage{newtxmath} 

\shorttitle{Large Scale Structure in Chiles}
\shortauthors{N. Luber et al.}

\begin{document}

\title{Large Scale Structure in CHILES using DisPerSE}
\author{Nicholas Luber\altaffilmark{1,2,3}}
\author{J. H. van Gorkom\altaffilmark{3}}
\author{Kelley M. Hess\altaffilmark{4,5}}
\author{D.J. Pisano\altaffilmark{1,2,6}}
\author{Ximena Fern\'andez\altaffilmark{7,8}}
\author{Emmanuel Momjian\altaffilmark{9}}
\email{nicholas.m.luber@gmail.com}

\altaffiltext{1}{Department of Physics and Astronomy, West Virginia University, P.O. Box 6315, Morgantown, WV 26506, USA}
\altaffiltext{2}{Center for Gravitational Waves and Cosmology, West Virginia University, Chestnut Ridge Research Building, Morgantown, WV 26505}
\altaffiltext{3}{Department of Astronomy, Columbia University, New York, NY 10027, USA}
\altaffiltext{4}{ASTRON, the Netherlands Institute for Radio Astronomy, Postbus 2, 7990 AA, Dwingeloo, The Netherlands}
\altaffiltext{5}{Kapteyn Astronomical Institute, University of Groningen, Landleven 12, 9747 AD, Groningen, The Netherlands}
\altaffiltext{6}{Adjunct Astronomer at Green Bank Observatory, Green Bank, WV, USA}
\altaffiltext{7}{Department of Physics and Astronomy, Rutgers, The State University of New Jersey, Piscataway, NJ 08854-8019, USA}
\altaffiltext{8}{NSF Astronomy and Astrophysics Postdoctoral Fellow}
\altaffiltext{9}{National Radio Astronomy Observatory, P.O. Box 0, Socorro, NM 87801, USA}

\begin{abstract}
We demonstrate that the Discrete Persistent Source Extractor (DisPerSE) can be used with spectroscopic redshifts to define the cosmic web and its distance to galaxies in small area deepfields.  Here we analyze the use of DisPerSE to identify structure in observational data. We apply DisPerSE to the distribution of galaxies in the COSMOS field and find the best parameters to identify filaments. We compile a catalog of 11500 spectroscopic redshifts from the Galaxy and Mass Assembly (GAMA) G10 data release. We analyze two-dimensional slices, extract filaments and calculate the distance for each galaxy to its nearest filament. We find that redder and more massive galaxies are closer to filaments. To study the growth of galaxies across cosmic time, and environment, we are carrying out an HI survey covering redshifts $z$ = 0 - 0.45, the COSMOS HI Large Extragalactic Survey (CHILES). In addition we present the predicted HI mass fraction as a function of distance to filaments for the spectroscopically known galaxies in CHILES. Lastly, we discuss  the cold gas morphology of a few individual galaxies and their positions with respect to the cosmic web. The identification of the cosmic web, and the ability of CHILES to study the resolved neutral hydrogen morphologies and kinematics of galaxies, will allow future studies of the properties of neutral hydrogen in different cosmic web environments across the redshift range $z$ = 0.1 - 0.45.
\end{abstract}


\section{Introduction}
\subsection{Cosmic Web}
\quad We have long known, both in observations and simulations, that the matter in the universe is situated in an intricate system of filaments, walls, voids, and clusters, appropriately referred to as the cosmic web. The first large-scale redshift surveys undertaken revealed large underdense regions with sharp boundaries traced by string-like filaments of galaxies \citep{1986ApJ...302L...1D}. Since then, numerous surveys have been conducted with large fields of view that probe fainter and fainter magnitudes. Results from surveys such as the Sloan Digital Sky Survey (SDSS) and the 6 Degree Field Galaxy survey (6dFGS) have provided detailed insights into the large-scale matter distributions in the local universe \citep{2000AJ....120.1579Y,  2004MNRAS.355..747J}. The SDSS has been used to map the distribution of galaxies, at low redshifts, with thorough completeness, and there have been successful implementations of many different methods to distinguish individual features of the cosmic web \citep{2009MNRAS.400..183K, 2014MNRAS.438.3465T}.

\quad The Cosmic Evolution Survey (COSMOS) is a galaxy survey with a much narrower field of view than 6dFGS, or SDSS that has been carried out to probe the properties of galaxies from the near universe out to redshifts of $z$ $\approx$ 6 \citep{2007ApJS..172....1S}. Both photometric and spectroscopic redshifts from the COSMOS data releases have been used extensively to identify the structures and the properties of galaxies in relation to these structures. Techniques to identify structures, such as adaptive smoothing, have been used to obtain scale-free measures of densities from which persistent structures can be extrapolated \citep{2007ApJS..172..150S}.  In this paper we explore another algorithm, the Discrete Persistence Source Extractor (DisPerSE), \citep{2011MNRAS.414..350S}, for use on a subfield of COSMOS. There is some difficulty in applying structure-identifying techniques to observational data sets due to the increasing incompleteness as higher redshifts are approached, and due to errors in the measurements. However, work has been done in the comparison of different algorithms' ability to identify the cosmic web in these large-scale observational data sets. \citet{2017arXiv170503021L} have recently explored these different structure identifying techniques and analyzed their abilities to recover mass distributions in the universe across different environments. They show the strengths of these different techniques in recovering different types of three dimensional structure and nicely summarize the ability of different algorithms and the different scientific goals they can help achieve. They found that DisPerSE is well able to extract the filamentary structure in simulated data and can recover volumes and masses of the different components of the cosmic web that are in good agreement with the corresponding predictions from independent simulations. Recently, DisPerSE has also been successfully used in observational data \citep{2017arXiv171002676K, 2017arXiv170208810L, 2017MNRAS.466.4692K}. This paper differs from the aforementioned papers because we are using a much smaller area than \citet{2017arXiv171002676K, 2017MNRAS.466.4692K}, and we use spectroscopic redshifts from GAMA as opposed to the photometric redshifts used in \citet{2017arXiv170208810L}. In CHILES we probe a limited redshift range with many galaxies at lower redshifts, where the uncertainties in photometric redshifts  produce relatively large uncertainties in the derived large scale structure.  

\quad While the dark matter properties of galaxies as a function of distance in the cosmic web have been well studied, with currently conflicting results,  studies of the baryonic properties as a function of distance from the cosmic web have started only recently \citep{2019MNRAS.483.2101G}. Specifically, to investigate how the gas content correlates to large-scale environment. Statistical studies of the local universe have shown a tight correlation between gas content and local galaxy density. Specifically galaxies in clusters are highly HI deficient. Detailed HI imaging studies of small samples of galaxies have identified possible mechanisms of gas accretion, and removal, in voids and clusters, respectively \citep{2012AJ....144...16K, 2017MNRAS.464..666B, 2009AJ....138.1741C}. Moreover, HI is being studied in galaxies residing on the filamentary network of the cosmic web, \citep{2017MNRAS.466.4692K}, as possible observational evidence to support the Cosmic Web Detachment (CWD) model \citep{2016arXiv160707881A}. The CWD model theorizes that when galaxies detach from the cosmic web, they are cut off from their supply of cold gas, and the efficient phase of star formation ends. In addition, the problem of the missing baryons can be explored by observations of the gas content of the cosmic web. Although these cold gas signals would be extremely weak, it is theorized that with higher resolution telescopes and longer integration times, the gas will become detectable \citep{2017arXiv170200193H}.

\subsection{CHILES}
\quad The COSMOS HI Large Extragalactic Survey, CHILES, is a deep spectral line radio survey of a 40 x 40 arcminute area of the COSMOS field, undertaken with the Karl G. Jansky Very Large Array (VLA), to study the atomic neutral hydrogen (HI) properties of galaxies across cosmic time. CHILES is the first survey to study the properties of neutral hydrogen in individual galaxies over a continuous redshift range from $z$ = 0 to $z$ = 0.45. The CHILES pilot, undertaken in 2011, covering a bit less than half the redshift range from $z$ = 0 to $z$ = 0.193 showed that with a single pointing, and 50+ hours of integration, that we could detect the atomic gas in galaxies at its highest redshift.  Thus demonstrating that the large data volume and abundant presence of man made radio frequency interference (RFI) could be adequately dealt with demonstrating that the VLA is ready to do a deep HI survey \citep{2013ApJ...770L..29F}.  At the completion of the 1000 hours of observation, the survey will provide HI masses, and resolved morphology and kinematics for over 300 galaxies in a variety of different environments, stellar masses, and redshifts. We have reduced, and imaged the first observing epoch, 180 hours. We succesfully characterized HI emission at the highest redshift to date in \citet{2016ApJ...824L...1F}, and we showed that we can derive meaningful results in the most challenging RFI environment in \citet{2019MNRAS.484.2234H}. The survey will also allow for statistical measures of HI properties as a function of redshift and various properties of galaxies via stacking, as well as provide for an estimate on the measurement of $\Omega$$\textsubscript{HI}$ as a function of redshift. In this paper, we test the use of DisPerSE, a topological algorithm that calculates density ridges in a distribution of points, which can be used to identify the large scale structure in the CHILES volume in order to measure the gaseous properties of the cosmic web.

\quad In Section 2 we discuss DisPerSE, the structure defining package on which this work is based. In Section 3, we discuss the development of the catalog we use throughout the rest of the paper in both the implementation of DisPerSE, and the subsequent scientific analysis, and how we slice up the data for further calculation and visualization. In Section 4, we present the testing we did on the parameters of DisPerSE and conclude what the best parameters are. In Section 5, we present as a sanity check results  on known  properties of galaxies as a function of distance from filaments in the cosmic web, and the predicted HI content for galaxies in the CHILES survey. Lastly we discuss a few examples of the HI morphology found for galaxies at different locations in the cosmic web.   In Section 6, we present our conclusions. Throughout this paper, we adopt  a flat $\Lambda$CDM cosmology with H\textsubscript{0} = 71 km s$^{-1}$ Mpc$^{-1}$, $\Omega\textsubscript{M}$ = 0.27, and $\Omega_{\Lambda}$ = 0.73.

\section{Methods}
\subsection{DisPerSE}
\quad We have chosen to use DisPerSE for defining the cosmic web because of its easy implementation on an observational dataset, and the fact that it is a scale free topological algorithm, which means that it can operate on different distance scales. There are many different ways, and algorithms that people use to define large scale structure. They are discussed at length in \citet{2017arXiv170503021L} where it is concluded out that DisPerSE sufficiently recovers the filaments of the cosmic web. DisPerSE, is an algorithm that uses discrete Morse theory to calculate large-scale structure from a distribution of points, with particular focus on extracting the cosmic web in either simulations or large redshift surveys \citep{2011MNRAS.414..350S}. DisPerSE relies on Morse theory, a mathematical framework in which Morse functions, a type of twice-differentiable smooth scalar functions, can be used to describe the topology of a region. This is possible due to the intricate relationship between a point distribution’s geometry and topology. The resulting description of the topology can then be used to define the structures in the point distributions. Discrete Morse theory is implemented as an alternative to Morse theory due to the fact that galaxy catalogs cannot be used to construct Morse functions with the necessary criterion. Instead, discrete Morse theory is an adaption that defines discrete functions over simplicial complexes allowing for discrete datasets, i.e. data from redshift surveys, to be properly analyzed.

\quad A Delaunay tessellation, a technique that outlines triangles over every point in a distribution such that the area of the triangle is inversely proportional to the density of the region where that point is located, is performed on the dataset and then used to calculate the density field using the Delaunay Tessellation Field Extractor (DTFE) \citep{2007PhDT.......433S}. Then, the discrete Morse-Smale complex, the intersection of the ascending and descending manifolds of discrete Morse-Smale functions, is derived from this density field to ascertain topological features with different indices corresponding to the various components of the cosmic web. The intersection is defined by a series of critical points, these points are where the gradients of the manifolds are equal to zero, and thus define the maxima or minima of that topological feature in the point distribution. The series of critical points that make up the discrete Morse-Smale complex can then be connected by a series of arcs to trace out the topological features. These features are visualized by returning the ascending 0, 1, 2 and 3 manifolds that correspond to voids, walls, filaments, and clusters, respectively.

\section{Catalog Development}
\subsection{Compilation}
\quad In order to provide for the most complete and accurate catalog of galaxies to utilize in the implementation of DisPerSE, we construct a catalog that combines data from different collaborations. The spectroscopic redshifts come from the GAMA G10 region, a GAMA target field coincident with COSMOS \citep{2015MNRAS.447.1014D}. The stellar properties come from the COSMOS 2015 photometric data release \citep{2016ApJS..224...24L}. GAMA is a collaboration that is undertaking deep spectroscopic redshift surveys in different small angular area fields. In particular, in the G10 region, GAMA is reprocessing zCOSMOS data to produce higher confidence spectroscopic redshifts \citep{2011MNRAS.413..971D}. The COSMOS region is a 2 \text{degree\textsuperscript{2}} area of the sky chosen for its lack of stellar density from our own galaxy, and the GAMA G10 region was chosen to be coincident with this region in order that GAMA may use the vast photometry available through COSMOS. The much smaller CHILES region is similarly coincident with the COSMOS region for the same reason of vast ancillary data. However, the CHILES region is slightly off center from COSMOS to avoid any  bright radio sources. We select every galaxy with a spectroscopic redshift from GAMA. This is done because the spectroscopic redshifts calculated by GAMA have been done with more sophisticated fitting techniques, and accuracy checks than what was done by zCOSMOS in their most recent release of spectroscopic redshifts \citep{2015MNRAS.447.1014D}. We select only those galaxies that GAMA has determined to have a high quality redshift determination, for which the mean uncertainty in redshift is 0.005. We then match these redshifts with the galaxies as they appear in the COSMOS 2015 catalog. We then, if available, add ancillary data from the COSMOS 2015 photometric catalog for each galaxy \citep{2016ApJS..224...24L}. This includes the stellar mass, the quiescent, or star-forming classification, and the K-corrected NUV-r color. The stellar mass is calculated using the BC03 template \citep{2003MNRAS.344.1000B}, in which the mass is derived from a fitted stellar spectrum. The quiescent and star forming classification was derived using the NUV-r/r-J relationship by the COSMOS team and the method is further elaborated in \citet{2016ApJS..224...24L}. There are a factor of ten more photometric redshifts from COSMOS than there are spectroscopic ones. However, due to the increased accuracy of spectroscopic redshifts, we find the loss of data to be an acceptable trade off. The final catalog that we synthesized contains 11500 galaxies in the COSMOS field, each with a redshift from GAMA, and stellar properties from COSMSOS2015.
  
\quad In Figure 1, we present a comparison between the contents of the photometric catalog compiled from COSMOS2015 photometry and the catalog compiled from GAMA redshifts over the COSMOS field. We see that the photometric catalog has significantly more coverage, with redshifts for roughly a factor 5 more galaxies. Especially at the lower redshifts the increased accuracy in using spectroscopic redshifts allows for a more accurate description of the large scale structure in the CHILES field. We show in the following sections that we are able to use DisPerSE to define the large scale structure using spectroscopic redshifts, despite the smaller number of galaxies with spectroscopic redshifts.  
  
\begin{figure*}  
\begin{center}
\includegraphics[scale = 0.5]{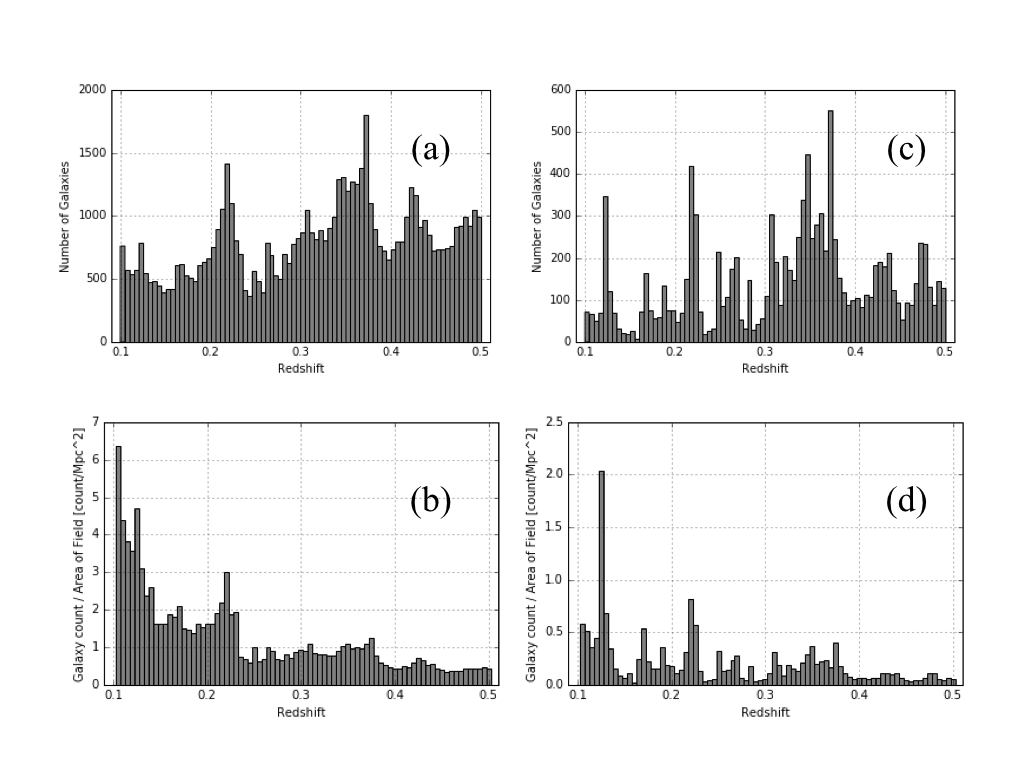}
\caption{Quantitative description of the photometric and spectroscopic galaxy catalogs across redshifts in bins of thickness $z$ = 0.01. \textit{a} The distribution of the population of galaxies with photometric redshifts. \textit{b} The number density of galaxies in physical units with photometric redshifts. \textit{c} The distribution of the population of galaxies with spectroscopic redshifts. \textit{d} The number density of galaxies in physical units with spectroscopic redshifts.}
\end{center}
\end{figure*}

\begin{figure}[h!]
\includegraphics[scale = 0.4]{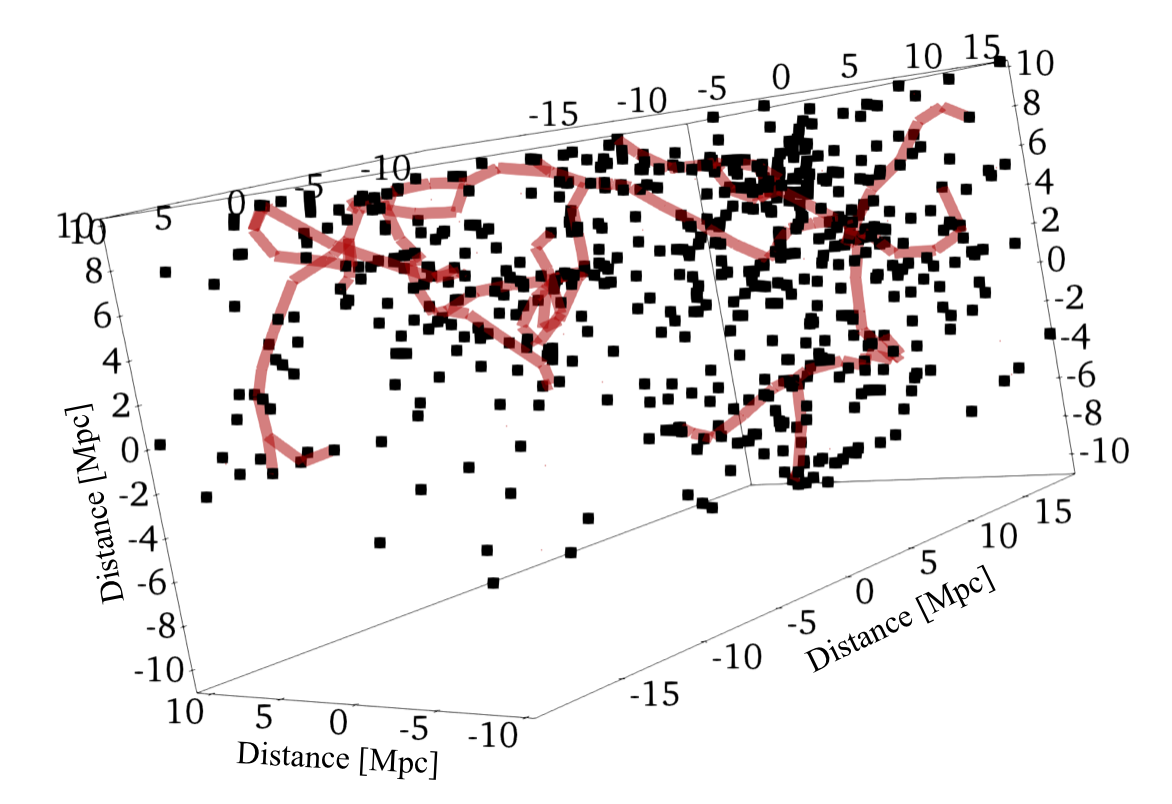}
\caption{An illustration of the three dimensional large scale structure that DisPerSE extracts in the redshift range $z$ = 0.21 - 0.22. Each point is at the location of a galaxy and the dark magenta lines indicate the extracted filaments. The axes are the physical size of the field, at each redshift, measured in Mpc.}
\end{figure}

\subsection{Two Dimensional Slicing}
\quad To assess the quality of DisPerSE in the extraction of the cosmic web, an examination by eye is necessary to confirm if the identified filaments trace the visual patterns in the images. However, it is impossible to determine the validity of the results due to our inability to visualize three dimensional structures on a two dimensional surface.  Although condensing the galaxies into a two-dimensional slice eliminates the calculation of physical distances from a galaxy to a filament, we are still able to calculate relative separation and maintain the ability to verify structure by eye. We use a redshift thickness of $z$ = 0.01 to ensure that we are not forcing galaxies to be part of a shared geometry that is too far separated in redshift. Furthermore, we  implement DisPerSE on these two dimensional slices, rather than isolating the best fit parameters and doing the analysis in three dimensions, for the aforementioned reason of visual analysis. Lastly, although this work is primarily concerned with applicability to CHILES, we implement DisPerSE over the full COSMOS region for two reasons. First, a larger sky area allows for topological structures to be more easily identified. Second, this allows for future work done with COSMOS data to use the large scale structure as defined by this paper. In future work, we plan to further test DisPerSE in three dimensional space.

\quad The use of spectroscopic redshifts in our study is a key difference in our definition of the cosmic web, as previous studies only use photometric redshifts.  An important consideration is the difference in relative uncertainty between photometric and spectroscopic redshifts. The mean uncertainty on a spectroscopic redshift is 0.005, while for our photometric redshift the mean uncertainty is 0.01. We are working in redshift slices of thickness of z = 0.01, and to use photometric redshifts, that have a similar uncertainty, would clearly cast a cloud of doubt over the calculated definition of the cosmic web. However, incompleteness of spectroscopic redshifts affects our results by not allowing us to accurately measure the density in areas that may be dominated by low surface brightness galaxies. Incompleteness in the spectroscopic data is a concern. The GAMA collaboration quotes a completeness down to 26 magnitudes. Figure 1D shows that at low to intermediate redshifts, the range probed by CHILES and future HI deepfields, the average number density of galaxies with spectroscopic galaxies goes down by only a factor 2, allowing us to be able to use spectroscopic redshifts across the entire range. For these reasons, we use spectroscopic redshifts in order to ensure that we recover meaningful structures, suitable for physical interpretation.

\quad In Figure 2, we illustrate the inaccessible nature of the visualization of a three-dimensional display of the cosmic web. Although there are clearly defined filaments in the image, it is impossible to distinguish by eye which galaxies do or do not belong to these filaments. For an example of a two-dimensional rendering of this same redshift slice with the same parameters, see Figure 5. In Figures 2, 3, 5, and 6 we choose to display the redshift slice 0.21 - 0.22 because of its variety of large scale structures while also lacking large over or under densities. We believe the results in this slice to be an accurate representation of DisPerSE's abilities. Additionally, it falls near the middle of the CHILES redshift coverage, which is important as it underscores the success of DisPerSE to describe the dataset that we, as the CHILES collaboration, are interested in. The slice shown is the entirety of the COSMOS field, because that is what was used to derive the filamentary network.

\section{Using DisPerSE}
\subsection{Boundary Type}
\quad One challenging aspect of defining structure in the COSMOS field is the relatively small field of view. At redshifts $z$ = 0.1 - 0.2 the field of view, in physical size, is between 5 and 10 Mpc. This small scale greatly affects the identification of structure near the boundaries, the physical edges of the field, while also having a significant impact on our understanding of the cosmic web throughout the entire field. There are four boundary types, referred to as \textit{btype}, possible in DisPerSE, and they treat how boundary particles are added to the galaxy catalog when running the internal DisPerSE function, \textit{delaunay\_2D}, to properly calculate the DTFE around the boundaries of the field. These four boundary types are described as follows: 

\textit{(i) Smooth:} particles are added to the boundaries based off of an interpolated density estimation

\textit{(ii) Void:} there are no boundary particles added.

\textit{(iii) Mirror:} the particles outside the field mirror the particles on the edge.

\textit{(iv) Periodic:} normal periodic conditions.

Smooth was eliminated as a choice for efficiency reasons. The Delaunay tessellation under that condition, after 10 hours, showed no sign of completion and added 20+ gigabytes worth of boundary points. This is in opposition to a run time of 10 seconds for the other boundary conditions.

\begin{figure*}[h!t!]
\begin{center}
\includegraphics[scale = 0.5]{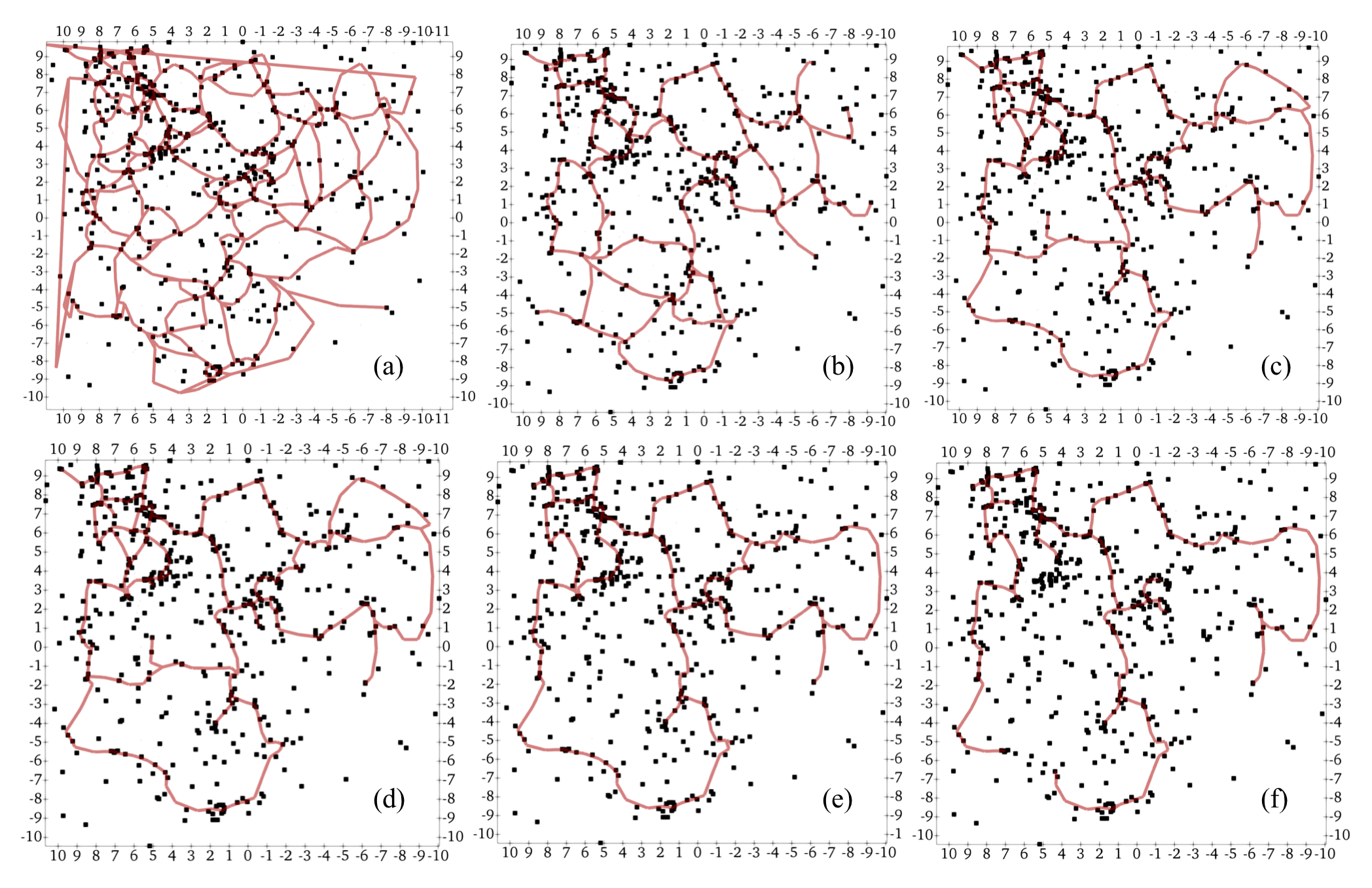}
\caption{The extracted filamentary network for the redshift range $z$ = 0.21 - 0.22 overlaid on the galaxy distribution. The axes are physical size of the field, at each redshift, measured in Mpc. \textit{a - c} The plots illustrate the effect that the following boundary types used for the Delaunay tessellation have on the resultant filaments: \textit{(a)} void, \textit{(b)} periodic, and \textit{(c)} mirror. Each of these plots has an \textit{nsig} of 1. \textit{d - f} The plots illustrate the effect that varying nsig has on the filaments. \textit{(d)} The filaments when nsig is equal to 1. \textit{(e)} The filaments when nsig is equal to 3. \textit{(f)} The filaments when nsig is equal to 5. In each of these plots we use a mirror boundary type.}
\end{center}
\end{figure*}

\quad In Figures 3a-c, we illustrate the  effect that a change in boundary has on the filaments that DisPerSE identifies in the same redshift slice. The void boundary condition clearly fails to return logical results on the boundaries of the field, with the filaments being jagged lines that connect galaxies on opposite ends of the field. Towards the center of the image, it is clear that too many spurious filaments are created. Toward the center of the image many lines are drawn, that seem not to follow a physical filament, but rather connect two regions of enhanced density. There is less noticeable difference between the effects that mirror and periodic boundary conditions have on the resultant filaments. However, we believe that mirror has extraneously extracted filaments in underdense regions. We also find that when periodic is used there are increased numbers of filaments in the overdense regions of the distribution, which are useful when designating walls of galaxies.  In our work we will use the periodic boundary condition. 

\subsection{Significance Level}
\quad  The boundary condition is not the only critical parameter in DisPerSE, at least equally important is the threshold that is used to decide whether structures are significant. In the main function of DisPerSE, \textit{mse}, there is a critical parameter referred to as \textit{nsig}. In \textit{mse} the user inputs the delaunay tessellation and the function returns the cosmic web. The parameter \textit{nsig} determines the persistence ratio threshold to use when determining whether a persistence pair is eliminated or not. A persistence ratio is the ratio of two points with consecutive indices in which a larger number indicates a stronger topological feature. A persistence pair are the two adjacent points on a topological structure used to derive the persistence ratio. Any persistence pair that has a persistence ratio less than the probability of the pair appearing in a random field will be canceled. Raising this parameter will eliminate extracted structures that may be spurious or less clearly defined by the galaxies in the field.

\quad In Figure 3d-f, we show the effects as the value for \textit{nsig} is varied over a galaxy distribution with all other parameters unchanged. As the parameter is raised, filaments that were previously extracted with a lower significance level, are not returned. After an examination of this effect over different redshift slices, we determined that the best value for nsig is 2. This is a result of the relative incompleteness of the survey at higher redshifts, which forces us to choose a lower value in order to identify any present structures. A lower value is also necessary to recover structures that may be better defined in three dimensions but that are harder to identify in two dimensions, such as a filament that is oriented mostly in the $z$-direction. Such a filament would have extension in a compressed plane and therefore the density ridges would not be as clearly defined in the two dimensional projection.

\quad Although the above selection process may be slightly subjective, we reached our conclusions by inspecting many different combinations of boundary conditions and thresholds. A brief summary of this selection process is simply put, that the significance level of 2 both returns reliable filaments while allowing for incompleteness at higher redshifts, and that periodic boundary conditions visually traces dense regions in the images. We decided to use a single value of nsig across the entire redshift range to simplify the comparison of different redshift slices. That way any analysis of trends across cosmic time are not obfuscated by a varying nsig parameter.

\bigskip
\bigskip
\subsection{Identifying Filaments}
\quad The ultimate goal of our use of DisPerSE  is to identify where a galaxy is located with respect to the cosmic web. We must find a way to determine the distance to a filament. Henceforth we define the cosmic web as the network of filaments, defined in DisPerSE by the ascending 2 manifolds. In order to understand a galaxy's placement in the cosmic web, we developed two methods to measure a galaxy's relative separation from the cosmic web.

\quad \textit{Critical Point Density:} In this method, we build a spheroid,in two dimensions this is a circle, and in three dimensions, a sphere, of a specified diameter, which grows larger as a function of redshift, around each critical point that comprises the filamentary structure. The higher the redshift, the larger the diameter to account for the greater distance between critical points. Then, for each galaxy, we add up how many of these spheroids the galaxy falls within. One strength of this method is that a threshold for the number of critical points near a galaxy can be further used to assess the large-scale structure. For example, a galaxy that is a neighbor of 10+ critical points is likely to be a member of a wall or a cluster. However, one weakness of this method is that it does not return any information about the physical distances between an individual galaxy and the cosmic web on the two-dimensional projection.

\quad \textit{Distance to Filament: } In this method, we take a galaxy and calculate the distance between the galaxy and the nearest critical point. By distance, we refer to the angular distance between a galaxy and the closest filament in the cosmic web. This is calculated using the scale factor as it would be for the central redshift of each slice. However, this method introduces an uncertainty in the distance measured, such that it assumes that a galaxy is closest to the cosmic web at the point of the nearest critical point, and not simply any point along the filament. Fortunately, this error is bounded by half the distance between the critical point used and the next closest one, which is a small distance, as the critical points well sample the filamentary network. The strength of this method is that despite the small errors it introduces, it produces meaningful distances between a galaxy and the cosmic web. One weakness is that the distances between galaxies and the cosmic web grow systematically larger at higher redshift, due to the fact that the critical points have larger separation. This is a result of increasing incompleteness in the COSMOS dataset at larger redshifts. The incompleteness creates lower spatial resolution in the output of DisPerSE. As the redshift slices become under-sampled, the filamentary network extracted is not indicative of the entirety of the cosmic web. The relative error on this measurement ranges from 0.1 - 5\%.

\quad In Figure 4 we present the distances of the galaxies with redshifts between 0.1 and 0.5 to filaments in the COSMOS field. We find a clear logarithmic relationship between number of galaxies and distance from the cosmic web. Moreover, considering that recent simulation work has shown that typical filament width varies from 0.1 - 1 Mpc, \citep{2016MNRAS.462..448G}, we calculate the number of galaxies that are within 1 Mpc of the cosmic web and find that 80\% of the galaxies in the volume are within these bounds. This is reasonable, insofar as we expect most galaxies to reside within the filamentary structures of the cosmic web, as opposed to being within voids.

\begin{figure}[h!]
\includegraphics[scale = 0.575]{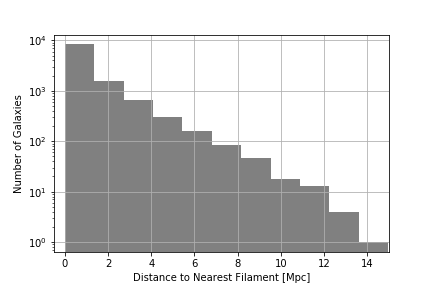}
\caption{A histogram of the galaxies in the redshift range $z$ = 0.1 - 0.5 as a function of distance from the cosmic web.}
\end{figure}

\quad When calculating distances we return both the critical point density and distance to filament for each galaxy. The combination of these will be used in future work to create further distinctions between different components of the cosmic web at different redshifts. We produce images that have both these quantities as attributes for each galaxy in the redshift slice. In Figure 5, we illustrate these two methods and the features they identify on the same dataset. The result of both methods show a distinction in different components of the cosmic web. The critical point density method, as seen in the upper panel of Figure 5, clearly highlights overdense regions, as opposed to the entire filamentary network. This is illustrated by the steep color gradient seen only in overdense regions, where multiple filaments meet. All other points are of almost uniform color. The distance method, as seen in the bottom panel of Figure 5, more clearly traces the individual filaments and properly illustrates the transition from filaments to the underdense regions. This is evident by the smooth color gradient from the filamentary network into the underdense regions.

\begin{figure}[h!]
\includegraphics[scale = 0.85]{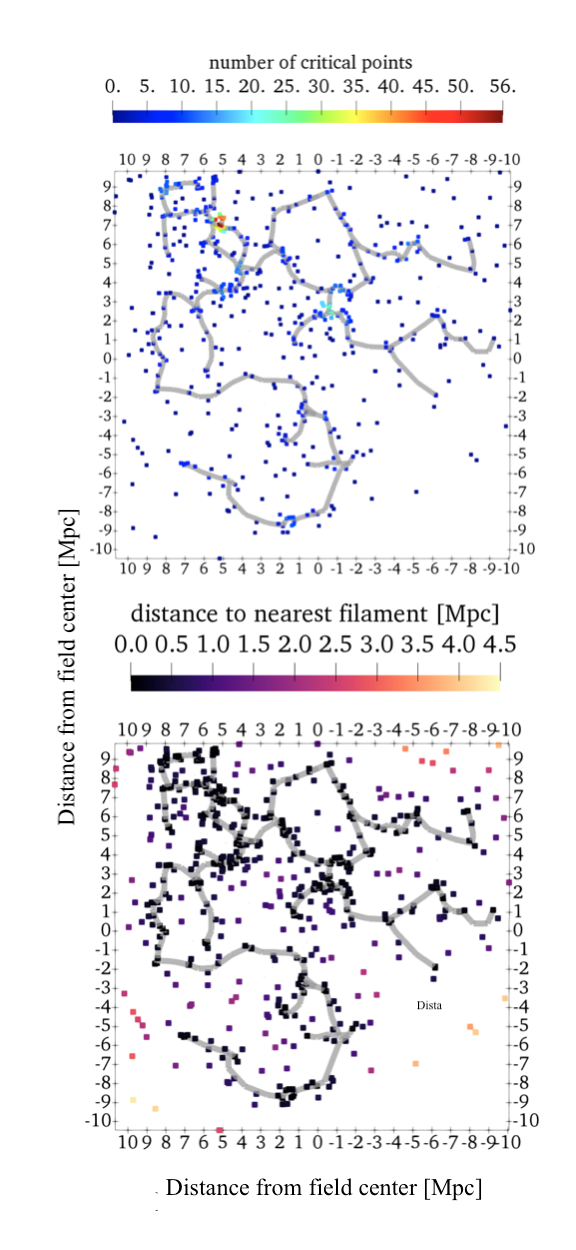}
\caption{The extracted filamentary network for the redshift range 0.21-0.22 overlaid on the galaxy distribution. \textit{Top: } The color scale for the galaxies corresponds to a number of neighboring critical points as described in the method, \textit{Number of Critical Points} in section 3.3. \textit{Bottom: } The color scale for the galaxies corresponds to the distance from the filament as measured by the method described in \textit{Distance to Filament} in section 4.3.}
\end{figure}

\subsection{Final Procedure Optimized for COSMOS}
\quad To identify the spine of the cosmic web, we use the following procedure:

(i) \textit{Split the Data:} We take our master galaxy catalog, provide lower and upper redshift, with a separation of $z$ = 0.01, and put  all galaxies in a given redshift bin a separate sub-catalog.

(ii) \textit{Use DisPerSE Methods:} We run a Delaunay tessellation on each redshift slice in the dataset with the boundary condition set to periodic. The output tessellation is then used as input in the \textit{mse} function, where we set nsig = 2, and return the filamentary network of the cosmic web in sky coordinates.

(iii) \textit{Convert Coordinates:} We then translate these original galaxy catalog and the returned filaments to the center of the COSMOS field and convert the coordinates into physical distances by using the scale factor for angular distance - physical distance that is determined by the median redshift in the slice.

(iv) \textit{Calculate a Galaxy's Position:} Then, we proceed to run the filament distance calculators and append both the number of critical points as well as the returned physical distances.

(v) \textit{Final Product:} We then return an image file for visualization for each galaxy that includes the log stellar mass, color, star forming classification, and cosmic web strength. 

\section{Applications}
\subsection{Procedure for Analysis}
\quad With the procedure for defining large scale structure defined in the previous sections, we now examine the physical properties of galaxies as a function of their placement within the large scale structure. Several large galaxy surveys have analyzed the observed properties of galaxies as a function of position within the cosmic web \citep{2016MNRAS.457.2287A, 2015ApJ...800..112G}. In this section we do the same for the galaxies in the COSMOS field to verify that our method provides sensible results, while also identifying any interesting trends. We also then discuss the predicted HI properties of the galaxies within the CHILES volume, and how they vary with placement within the cosmic web. 

\quad In order to display relationships with distance to the cosmic web, we had to present the distance axis in such a way that it reflects the changing scales of the extracted filaments, and the increased distance between the critical points that define the cosmic web. This increased distance between critical points is a result of incompleteness increasing at higher redshifts. The filamentary network is sampled by the same number of critical points, but their intrinsic spacing increases as the physical field of view increases with redshift. This is as a result of the lower number of galaxies due to the magnitude limits of a survey. We address this by normalizing each distance by the average separation of all the critical points and their nearest neighbors. This allows all of the galaxies' relative distances to be compared to one another in a physically more meaningful way.

\quad When analyzing the cosmic web across the whole redshift range, we proceed along the following method. We run DisPerSE, with all of the preceding specifications, and then calculate the distance of a galaxy to the filament, as described in Section 4.3. However, we only save the filament distance data for the galaxies in the inner half of the redshift range, for example, for the redshift slice $z$ = 0.21-0.22 , we only append the galaxies with $z$ $\geq$ 0.2125 and $z$ $<$ 0.2175. We do this because the geometry of galaxies at the edge of the slice may be more effected by the distribution of galaxies that are not included within the slice. For example, a galaxy at redshift 0.2105 will be more affected by the galaxies at 0.2095 than those at 0.2175. The incrementing of slices, gives a better description of the geometry around a galaxy. 

\begin{figure}[h!]
\includegraphics[scale = 0.85]{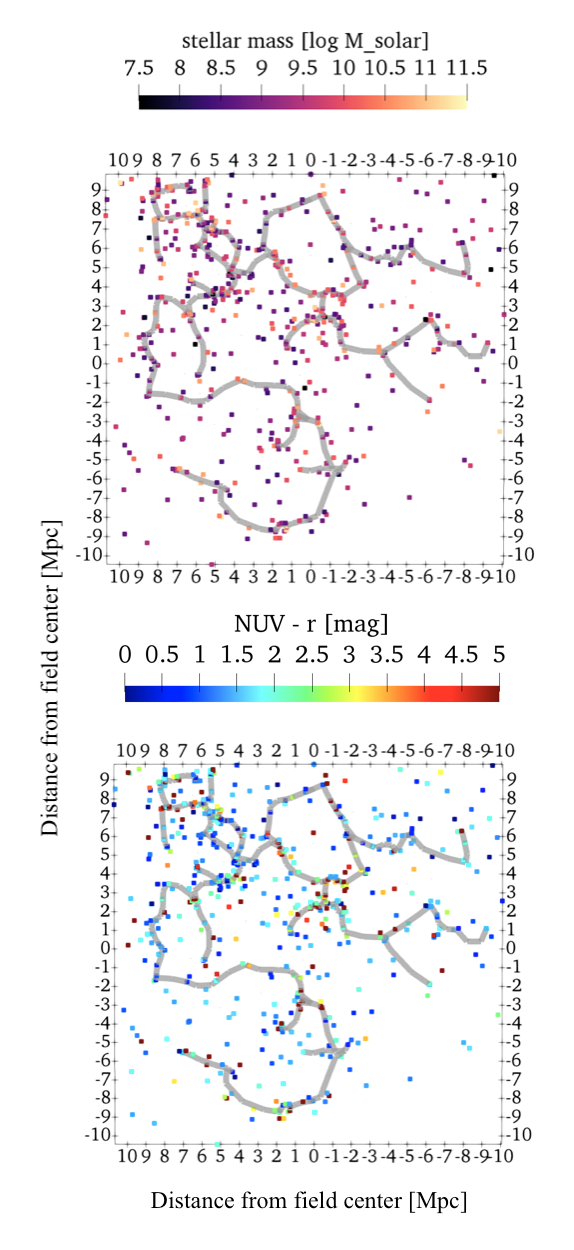}
\caption{The extracted filamentary network for the redshift range 0.21-0.22 overlaid on the galaxy distribution. \textit{Top: } The color scale for the galaxies corresponds to the log stellar mass of each galaxy. \textit{Bottom: } The color scale for the galaxies corresponds to the NUV-r color for each galaxy.}
\end{figure}

\begin{figure}[h!t!]
\includegraphics[scale = 0.65]{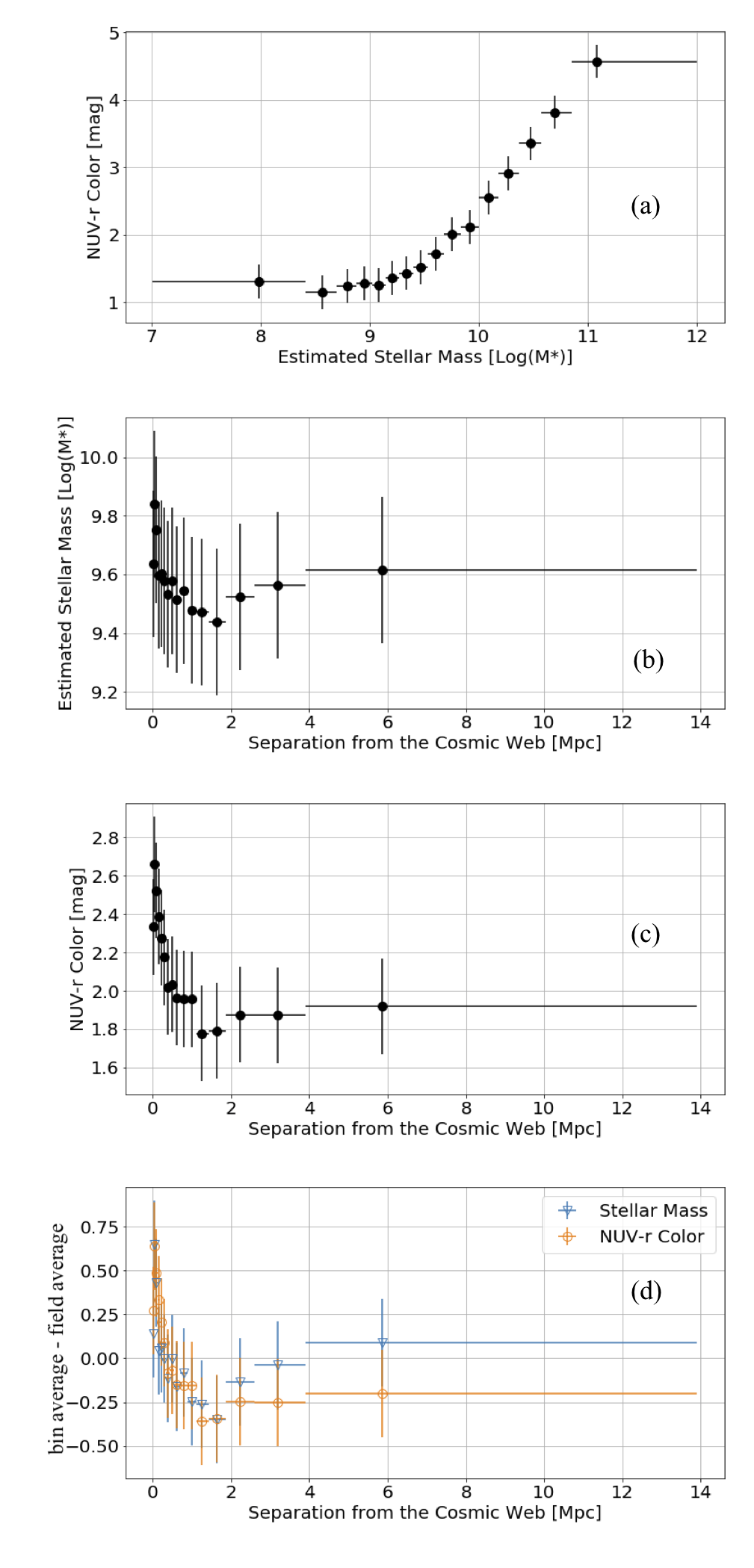}
\caption{\textit{(a)} Here we illustrate the Color vs. Mass relationship by sorting the galaxies by stellar mass, bining them into equal bins, and then finding the average NUV-r color and log stellar mass. \textit{(b - c)} Here we sorted the galaxies by separation from the cosmic web, binned them in bins of equal number of galaxies, and found in, \textit{(b)}, the average log stellar mass, and in, \textit{c}, the average NUV-r color. \textit{(d)} Here we rescaled the plots in b and c and superimposed them.}
\end{figure}

\bigskip
\subsection{Color and Mass versus Environment}
\quad We will now show some results on the optically derived properties of spectroscopically known galaxies in COSMOS as a function of location in the cosmic web, derived using DisPerSE.  

\quad \textit{Color vs. Stellar Mass} Figures 7a shows the color - mass relationship for the COSMOS field galaxies. The plot shows a strong correlation between the average of these two properties and it also shows that above a log stellar mass of 9.5 the redder galaxies dominate the densities of the bins, thus causing the clear upward trend. Similar analysis of the properties of galaxies in large redshift surveys have shown this. For example, the  VIMOS Public Extragalactic Redshift Survey (VIPERS) studied massive galaxies with intermediate redshifts and found a similar significant trend indicating that galaxies with higher masses are, on average, redder than the lower mass galaxies \citep{2017A&A...602A..15C}.

\quad \textit{Distance vs. Stellar Mass} Figure 7b shows the relation between stellar mass and distance from the cosmic web.  The average mass of galaxies decreases as the distance of the galaxy from the cosmic web increases from 0 - 1.5 Mpc. However, after a separation of 1.5 Mpc, a galaxy's average mass plateaus. Similar results are shown in a variety of low to intermediate redshifts in SDSS, where, independent of local density of galaxies, the galaxies along density ridges or filaments, are on average, more massive \citep{2017MNRAS.466.1880C}. Another illustration of the effects that the large-scale environment has is shown in \citet{2017ApJ...846L...4R}, where stellar mass, elliptical to spiral ratio and star formation rate are examined for galaxies as a function of distance from the center of a void. They find that as distance increases from the center of a void the elliptical to spiral ratio  and mass increase, while the star formation rate decreases. In  \citet{2017arXiv171002676K}, a similar trend in mass and color as a function of distance from the cosmic web is found  at lower redshifts, $z$ $\approx$ 0.2 in the G12 region.  

\begin{figure*}
\begin{center}
\includegraphics[scale = 0.5]{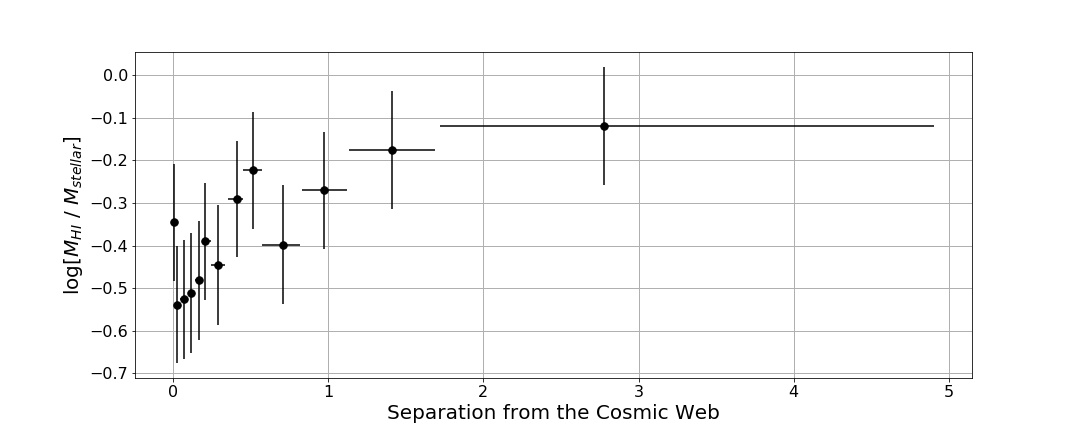}
\caption{Here we present the relationship between neutral hydrogen to stellar mass ratio and separation from the cosmic web. Each point represents the average distance from the cosmic web for 28 galaxies that are considered detectable by CHILES.}
\end{center}
\end{figure*}

\quad \textit{Distance vs. Color} Figures 7c illustrates a correlation between a galaxy's color and its distance from the cosmic web. The effect of distance on log NUV-r color is similar to the effect distance has on log stellar mass. We see a turn at a distance of 1 Mpc from a gradual approach towards bluer galaxies to a plateau. This turn in slope indicates that the distance bins are no longer being affected by redder galaxies on the same scale that they were before the turn. Moreover, the gradual slope before 1 Mpc indicates a smooth decrease in the fraction of red galaxies as the distance from the cosmic web increases. Similar results have been seen in filaments leading to a cluster environment, where it was shown that the fraction of blue to red galaxies increases dramatically as the distance to the filament increases \citep{2007A&A...470..425B}. 

\quad In Figure 7d we superimpose plots 7b and 7c with the scales changed to a relative scale. We do this to illustrate the similarities between the trends in the two properties. Both have a similar drop off in slope between 0 - 1.5 Mpc from the cosmic web and then both have a similar upturn at 2 Mpc from the cosmic web. The mechanism causing this upkick is unclear, but the fact that we see it in both properties could indicate either an observational bias, or a physical mechanism at work.

\quad In summary, a synthesis of previous work, and our work with DisPerSE and the COSMOS dataset, indicates that redder galaxies are more massive and have less separation from the cosmic web while blue galaxies tend toward the less massive side and are observed across both filaments and underdensities. However, this statement relies on averages of galaxy properties and the detailed study of individual galaxies and the cosmic web environment is crucial to understand the role large-scale structure plays in galaxy evolution. In the future we will track these properties across cosmic time. 

\quad In Figure 6, we present images of the cosmic web at a redshift of $z$ = 0.21 - 0.22 in order to display a typical output of our method for any redshift slice. The figure shows that the redder and more massive galaxies are closer to the filament.

\subsection{Preliminary HI Results}
\quad The first 178 hours of CHILES data have been reduced, and imaged. In these first 178 hours, we have detected and analyzed the properties of 16 galaxies at redshifts $z$ $\approx$ 0.12 and $z$ $\approx$ 0.17, a redshift range which is strongly affected by RFI to demonstrate what we will be able to do with the data \citep{2019MNRAS.484.2234H}. In  \citet{2019MNRAS.484.2234H}, we identified the cosmic web in the procedure outlined in this paper, and examined their properties with respect to the cosmic web. The majority of the HI detections appear to lie within 100 kpc of a filament. Some of the detections show interesting alignment of the HI distribution of the galaxies and the orientation of the filaments. The most massive group of five galaxies is in the vicinity of an intersection of filaments, which can be characterized as a wall  of galaxies. This is in good agreement with  cosmological simulations that predict the most massive galaxies to be in dense environments.  The two most isolated galaxies have very extended HI and a relatively low star formation rate, in good agreement with what has been found  in surveys of void galaxies \citep{2012AJ....144...16K}

\quad The CHILES collaboration recently published the detection of neutral hydrogen in a star-bursting galaxy at redshift $z$ = 0.376, COSMOS J100054.83+023126.2, the highest redshift HI emission detection to date \citep{2016ApJ...824L...1F}. Figure 1b of \citet{2016ApJ...824L...1F} shows that the HI morphology is asymmetric within the galaxy and very extended to the south. This extension is aligned with the cosmic web as identified by DisPerSE. The galaxy lies on a North-South filament and the gas distribution is reflective of this. Additionally, the high concentration of gas to the south of the galaxy is pointing towards an intersection of filaments and high density region of galaxies perhaps indicative of some interaction between the detection and the wall.

\quad The small sample of galaxies currently detected in CHILES, the 16 galaxies in \citet{2019MNRAS.484.2234H}, and the 1 in \citet{2016ApJ...824L...1F}, is too small to provide significant  insights, but as HI is detected in more galaxies across cosmic time, these observations of HI morphology and placement in the cosmic web could allow for a deeper understanding of the evolution of galaxies. Further work will be done to characterize HI morphology quantitatively to more deeply analyze the effects of large-scale structures on HI perturbation and galaxy evolution.

\quad In Figure 8 we show what we can expect to find in the CHILES survey. It shows the average neutral hydrogen mass to stellar mass ratio of galaxies detectable by CHILES as a function of separation from the cosmic web. For this Figure we calculated photometric gas fractions taking galaxies that have both SDSS spectra and GALEX colors, calculating stellar masses using \citet{2003ApJS..149..289B} and g,r, and i magnitudes from SDSS. We then adapted results from \citet{2010MNRAS.403..683C} by taking a color cut at $NUV-3=3.4$ magnitudes and assuming an H~I to stellar mass fraction of 28\% for blue galaxies and 1.6\% for red galaxies. We use a 5$\sigma$ detection limit which is calculated assuming a profile width of 150 km s$^{-1}$, a 1000 hours on source, and perfect gaussian noise. We only consider galaxies within the full width half power of the primary beam and we correct for the primary beam response. With this criterion we consider 364 galaxies to be detectable by CHILES, and are therefore used in the creation of this plot. We show that we expect to find that on average, as galaxies become farther from the cosmic web, their HI to stellar mass ratio increases. Local surveys have found conflicting results on this (see for example \citet{2017MNRAS.466.4692K},and \citet{2016ApJ...824..110O}) 

\section{Conclusions}
\quad We have shown that DisPerSE can be utilized to define the cosmic web in deepfields of a small angular area with the use of spectroscopic redshifts. This will be useful for the many upcoming small area HI deepfields, being carried out by MeerKat, GMRT, and ASKAP.

\quad Knowing the details of the cosmic web has many implications for our understanding of galaxies and galaxy evolution. We have developed a method to identify the cosmic web that utilizes DisPerSE and from this produced a schematic of the large scale environment of the COSMOS field across the CHILES redshift range. A few choice observations on some trends in the COSMOS data were presented. We demonstrated the correlation between a galaxy's color, stellar mass, and displacement from the cosmic web. The correlation of red, high mass galaxies with low cosmic web displacement and blue, low mass galaxies with high cosmic web displacement may begin to be further investigated, and its role in galaxy evolution understood. These results are generalized and only undertaken over a small field of view but their implications in the role of large scale structure in galaxy evolution are critical in directing further investigations to tackle more specific issues.

\quad As the remaining CHILES data is processed and imaged, we plan to use the filaments extracted with DisPerSE to understand the dependence of neutral hydrogen on large scale structure. We will examine morphologies, kinematics, and HI mass to stellar mass ratio in the different environments and identify any dependence on large scale structure for these properties. At the completion of our 1000 hours of observation we will be able to identify statistical properties of HI in galaxies that are in different components of the cosmic web at redshifts that have never before been probed. These resolved high redshifts detections will allow for key insights in the role of large scale structures in galaxy evolution. Furthermore, through the efforts of stacking and image smoothing we will attempt to identify cold gas in the filaments in between galaxies, a vital step necessary to understand the role of gas and the cosmic web in the evolution of galaxies.

\acknowledgments
\quad We acknowledge useful discussions with the entire CHILES collaboration. This work was in part supported by the National Science Foundation under grant number AST-1413102 to Columbia University. DJP acknowledges partial support from National Science Foundation under grant number AST-1412578.

\quad The National Radio Astronomy Observatory is a facility of the National Science Foundation operated under cooperative agreement by Associated Universities, Inc. X.F. is supported by an NSF-AAPF under award AST-1501342.

\bibliographystyle{yahapj}

\end{document}